# Novel imaging revealing inner dynamics for cardiovascular waveform analysis via unsupervised manifold learning

Shen-Chih Wang: M.D., Ph.D.

1. Department of Anesthesiology, Taipei Veteran General Hospital, Taipei, Taiwan
2. National Yang-Ming University, Taipei, Taiwan

Hau-Tieng Wu: M.D., Ph.D.

1. Department of Mathematics, Duke University, Durham, NC; Department of Statistical Science, Duke University, Durham, NC
2. Mathematics Division, National Center for Theoretical Sciences, Taipei, Taiwan

Po-Hsun Huang: M.D., Ph.D.

1. Department of Critical Care Medicine, Taipei Veterans General Hospital, Taipei, Taiwan
2. Cardiovascular Research Center, National Yang-Ming University, Taipei, Taiwan

Cheng-Hsi Chang: M.D.

1. Department of Anesthesiology, Shin Kong Wu Ho-Su Memorial Hospital, Taipei, Taiwan
2. College of Medicine, Fu Jen Catholic University, New Taipei, Taiwan

Chien-Kun Ting: M.D., Ph.D.

1. Department of Anesthesiology, Taipei Veteran General Hospital, Taipei, Taiwan
2. National Yang-Ming University, Taipei, Taiwan

Yu-Ting Lin: M.D., Ph.D.

1. Department of Anesthesiology, Taipei Veteran General Hospital, Taipei, Taiwan

**Address for correspondence**: Yu-Ting Lin, Attending Anesthesiologist, Department of Anesthesiology, Taipei Veterans General Hospital. 201, Section 2, Shih-Pai Road, Taipei, Taiwan. Telephone number: 886-2-28757549, Fax number: 886-2-28751597, E-mail: linyuting@hotmail.com.tw

# Abstract

**Background:** Cardiovascular waveforms contain information for clinical diagnosis. By "learning" and organizing the subtle change of waveform morphology from large amounts of raw waveform data, unsupervised manifold learning helps delineate a high-dimensional structure and display it as a novel three-dimensional (3D) image. We hypothesize that the shape of this structure conveys clinically relevant inner dynamics information.

**Methods:** To validate this hypothesis, we investigate the electrocardiography (ECG) waveform for ischemic heart disease and arterial blood pressure (ABP) waveform in dynamic vasoactive episodes. We model each beat or pulse to be a point lying on a manifold, like a surface, and use the diffusion map (DMap) to establish the relationship among those pulses. The output of the DMap is converted to a 3D image for visualization. For ECG datasets, first we analyzed the non-ST-elevation ECG waveform distribution from unstable angina to healthy control in the 3D image, and we investigated intraoperative ST-elevation ECG waveforms to show the dynamic ECG waveform changes. For ABP datasets, we analyzed waveforms collected under endotracheal intubation and administration of vasodilator. To quantify the dynamic separation, we applied the support vector machine (SVM) analysis and reported the total accuracy and macro F1 score. We further carried out the trajectory analysis and derived the moving direction of successive beats (or pulses) as vectors in the high-dimensional space.

**Results:** For the non-ST-elevation ECG, a hierarchical tree structure comprising consecutive ECG waveforms spanning from unstable angina to healthy control is presented in the 3D image (accuracy = 97.6%, macro-F1=96.1%). The DMap helps quantify and visualize the evolving direction of intraoperative ST-elevation myocardial episode in a 1-hour period (accuracy=97.58%, macro-F1=96.06%). The ABP waveform analysis of Nicardipine administration shows inter-individual difference (accuracy=95.01%, macro-F1=96.9%) and their common directions from intra-individual moving trajectories. The dynamic change of the ABP waveform during endotracheal intubation shows a loop-like trajectory structure, which can be further divided using the manifold learning knowledge obtained from Nicardipine.

**Conclusions:** The DMap and the generated 3D image of ECG or ABP waveforms provides clinically relevant inner dynamics information. It provides clues of acute coronary syndrome diagnosis, shows clinical course in myocardial ischemic episode and reveals underneath physiological mechanism under stress or vasodilators.

# Introduction

Cardiovascular waveforms contain abundant information for clinical diagnosis. The pulsatile waveform morphology reflects physiological dynamic interaction inside the human body. Traditionally, the waveform morphology knowledge is condensed as features derived from these waveforms. For example, the QRS complex and the ST segment in electrocardiography (ECG) help in the diagnosis of arrhythmia and ischemic heart disease[1-3]. Augmentation index and other features of arterial blood pressure (ABP) waveform analysis have been applied to assess the medical treatment of chronic hypertension or to assess the cardiovascular status.[4-7]

However, these derived features may not represent all available information in waveforms, and we need a new tool to grasp more information for inner dynamic physiological interactions in some clinical situations. Manifold learning is a recently developed data analysis technique that extracts inner structure from the data. It belongs to the family of unsupervised machine learning methods.[8] From the input of waveform data, it delineates a representative geometric structure in a high-dimensional space, referred to as a manifold.[9,10] The diffusion map (DMap) is one of the manifold learning algorithms undergoing rapid development in recent years.[10] We hypothesize that manifold learning might be useful to depict a complicated physiological dynamic, and further display them in a three-dimensional (3D) "image" to gain a comprehensive picture of the underlying dynamical physiological process. The novelty of this approach is that it depicts the inner dynamics directly from the raw waveform, rather than the derived features. To the best of our knowledge, this is the first study applying manifold learning to cardiovascular waveform analysis.

In this study, we explore cardiovascular waveforms by the DMap, and convert them into 3D images. We show that the pattern in the 3D image is possibly relevant to clinical conditions. Specifically, ECG waveforms with respect to ischemic heart disease were investigated with two conditions, ST-elevation myocardial infarction and non-ST elevation myocardial infarction. We also investigated two ABP waveform datasets during antihypertensives administration and endotracheal intubation. We also provide statistical quantification of the visualized patterns.

# Methods

## Subjects

We analyzed two ECG datasets and two ABP datasets from prospective observational studies and public databases. The DMap provides a novel representation of the waveform. We used it to construct a 3D image as a guide for further analysis to quantify information relevant to clinical condition. (Fig. 1)

To investigate the ECG waveform in Non-ST-elevation myocardial infarction (NSTEMI), we collected first dataset from the Massachusetts General Hospital-Marquette Foundation (MGH/MF) Waveform database with cardiovascular waveforms of 250 critical patients from PhysioNet[11,12]. From this database, two patient groups were collected: the unstable angina group and ischemic ECG group. The unstable angina group comprises patients diagnosed as unstable angina or angina in the pertinent history. The ischemic ECG group comprises patients whose ECG was interpreted as ischemia or subendocardial injury without unstable angina in their current diagnosis. We excluded patients with a pacemaker or bundle branch block. We selected 10-min consecutive ECG beats from the beginning of data recording. We omitted those ECG segments with ectopic beat, motion artifact, or spurious baseline wandering.

The second dataset, Intra-operative ST-Elevation (ISTE) dataset, is composed of an ECG signal recorded during a coronary artery bypass grafting surgery during which an episode of ECG waveform changes accompanied with a decrease in heart contractility. The lead II ECG waveform appeared to have a rapid ST-elevation, suggesting possible inferior wall myocardial ischemia (Supplemental Figure S2). During this episode, the surgeons and the anesthesiologists also noticed a decrease in heart contractility by direct vision and trans-esophageal echocardiography. After this ST-elevation event, the heart contractility improved, and the abnormal ECG waveform ameliorated gradually. The patient recovered uneventfully. We analyzed a 1-min segment of ECG starting 20 s before the episode of ST-segment abnormality as a demonstration of the dynamical waveform morphology. We also extended the observation period of the above ISTE data to a 66-min segment of ECG starting 1-min before the episode to analyze the ECG waveform evolving with time. For a comparison, we added a 10-min ECG signal from another case with a good functional status (American Society of Anesthesiologists physical status classification II) undergoing simple laparoscopic surgery as a healthy control.

The third dataset, Nicardipine dataset, contains the ABP waveform recorded during intravenous bolus dose administration (1 mg) of Nicardipine, a calcium channel blocker, for high blood pressure during surgery. This dataset includes 12 cases. Each case started from the time the bolus dose was given and ended when the blood pressure no longer monotonically decreased. These 12 cases range between 100 to 300 s and contribute 2364 pulses in total.

The fourth dataset, Endotracheal Intubation (ETI) dataset, contains 9 ABP waveform data recorded during the endo-tracheal intubation in the beginning of the routine general anesthesia. Each case started 1 min before the procedure and lasted for 200s, comprising 223 pulses. In total there are 2,957 pulses.

We also combined the Nicardipine dataset and ETI data set for further analysis: partitioning the ETI dataset according to the knowledge learned from the Nicardipine dataset.

Except the MGH/MF dataset, all datasets were collected prospectively from patients undergoing general anesthesia and surgery for medical reasons. The ISTE and ETI datasets collection were approved by the institutional ethics review board of Shin Kong Wu Ho-Su Memorial Hospital, Taipei, Taiwan (IRB No.: 20160106R), and the Nicardipine dataset collection was approved by Taipei Veterans General Hospital, Taipei, Taiwan (IRB No.: 2017-12-003CC). Written informed consent was obtained from each patient. Physiological waveforms of the ISTE and ETI datasets were collected from the standard patient monitoring, Philips IntelliVue™MP60 and MX800 (Philips Healthcare, Andover, Massachusetts, USA) via a third-party software, ixTrend™ Express ver. 2.1 (ixitos GmbH, Berlin, Germany). The sampling rates of ECG (lead II in the EASI mode)[13] and ABP were 500 Hz and 125 Hz, respectively. Physiological signals of the Nicardipine dataset were collected from another standard patient monitoring, GE CARESCAPE™B850 (GE Healthcare, Chicago, Illinois, USA) via the data collection software, S5 collect (GE Healthcare, Chicago, Illinois, United States). The sampling rates of all waveforms were 300 Hz.

All anesthetic managements for each patient in our observational studies were performed per institutional standard practice and under the discretion of the anesthesiologist. The ABP was obtained directly from the percutaneous radial arterial cannulation with a continuous pressure transduction.

**Manifold learning and diffusion map**

The idea of manifold learning is hypothesizing that the dataset lies on a curve or surface, or more generally a manifold. A manifold is a geometric structure that is an extension of the 3D solid object as many commonly encountered items in our daily life to the high dimensional space. In this work, we model each beat or pulse in a waveform to be a point lying on a manifold.[10] For example, for an ECG signal with 100 cardiac cycles, we have 100 points lying on a manifold. Usually, the dimension of a point is high, and directly deriving information, or feature, out of the dataset is critical and nontrivial.

With this model in mind, we use the manifold learning tool, the DMap, to establish the relationship among those points and to derive the inner dynamic features as clinical information. The output of the DMap can be converted to a 3D image for visualization. In the 3D image, each pulse is represented as a new point, and all the pulse waveforms are represented as a point cloud. While the 3D image is useful for visual inspection, it only contains partial information of the DMap output. The diffusion distance (DDist)[10] represents the similarity between two pulses by their straight-line distance in the output of DMap. With the DMap, on the microscopic scale, a smaller DDist indicates the more similarity between two pulses; on the macroscopic scale, the network-like inter-relationships among all pulses form the shape that can be seen as a 3D image. From the machine learning perspective, the DMap is a nonlinear dimensional reduction method[9], which is a generalization or a counterpart of the commonly applied linear dimensional reduction method, principal component analysis.[14] In this study, the result of the DMap is visualized by using 3D imaging, and the quantification is based on the DDist. More technical information is presented in Supplementary and the references therein.

**Waveform Data Analysis**

We used the R peak location of each ECG cycle or the maximum of the first derivation of each ABP cycle during the ascent as a fiducial point. To obtain pulses or beats from the waveform, an adequate time window was chosen to cover one cycle beat (or pulse). As the duration of each beat (or pulse) was not constant, we truncated those clean beats to be of a uniform size according to their minimal lengths. In the MGH/MF dataset, we rescaled all ECG waveforms to be aligned with P wave, R peak and T to normalize the QT interval inter-individual variations. The rescaling was performed on each waveform cycle with linear interpolation oversampling to 2,000 points and then downsampling to 500 points per cycle. Only MGH/MF dataset underwent this rescaling procedure. Then, we normalized each pulse by removing the mean and setting the variance of each pulse to 1 so that only the morphology information remains. The collected normalized pulses are input to the DMap to generate the 3D images for the upcoming analysis.

The similarity between two groups was determined as the DDist between their geometric centers, which were the mean positions of all points within their groups. For the trajectory analysis, we derived the moving direction of successive pulses as vectors in the high-dimensional space; technically speaking, it is the slope of the trajectory in the high-dimensional space. The step-wise-included angles between the two trajectories are calculated by the arccosine function. It is worth mentioning the meaning of the angle measurement. Two perpendicular trajectories indicate low relationship between each other, whereas two parallel trajectories, either in the same or opposite direction, mean a high relationship. It is a principal commonly used in data science.

**Statistical analysis**

As the output of the DMap is high-dimensional, we use the support vector machine (SVM), a multivariate classification tool[15], to quantify the dynamics separation. The SVM is a supervised machine learning algorithm. We performed the (multiclass-) SVM with the polynomial kernel function. In the multiclass-SVM, the one-against-one policy and the default parameters (Gamma=1.0, Degree=3, Coef0=0) without parameter tuning. To further avoid overfitting, we performed a four-fold cross validation, where for each database, we construct four folds by dividing each group into four subsets consisting of consecutive pulses. To handle the imbalanced data, we applied the SMOTE (Synthetic Minority Over-Sampling Technique)[16] to the training dataset. A group was considered imbalanced if it contained less than 15% points of the largest group. The means of total accuracies and macro-F1's of the four-fold CV were reported (The confusion matrix, the sensitivity and precision of each group, and a sensitivity analysis are shown in Supplementary C.1). The F1 of a group is the harmonic mean of that group's sensitivity and precision, and the macro-F1 is the mean of F1's of all groups. Descriptive data were presented as median (interquartile range). The DDist between two groups was expressed as the mean and 95% confidence interval, which were calculated by the bootstrap resampling without replacement in 100,000 samples[3,13]. The hypothesis test for the scalar variable data, such as the comparison of DDist's of two groups, was two tailed, performed by the bootstrap resampling[3,13] without replacement in 100,000 samples. A p value <0.05 was considered statistically significant. The waveform 3D visualization was performed using Visual Studio® community 2019 (Microsoft, Redmond, Washington, USA). The SVM was performed with the standard libsvm library[17], and all statistical analyses were performed using the standard statistical program R (version 3.6.1).

## Results

**NSTEMI ECG Dataset**

There are 10,583 ECG beats collected from the MGH/MF database for analysis, including 1,612 (2,200s) ECG beats from three unstable angina cases, 8,247 beats from 12 ischemic cases, and 724 ECG beats from the healthy control. The health control is the same as that in the ISTE data. In this database, we do not consider the dynamics but simply the clustering effect of different groups.

The 3D image constructed from the DMap of the ECG beats spanning from the start of the P wave to the end of T wave is shown in Figure 1 (and supplemental video F1). This shows clusters with respect to the three groups. It also shows an aggregation of the unstable angina group and the ischemic-ECG group, whereas the healthy control is relatively isolated. Quantitative measurements show that the inter-group DDist between the unstable angina group and the ischemic ECG group is 0.170 (0.165, 0.176), which is significantly shorter than that between the ischemic ECG group and the healthy control as 0.418 (0.413, 0.422), and the DDist between the unstable angina group and the healthy control is 0.484 (0.479, 0.489). Both hypothesis tests show statistically significant differences ($p<10^{-5}$). The SVM analysis shows a total accuracy of 97.6% and macro-F1 96.1%.

**ISTE ECG Dataset**

The ISTE data represent the ECG waveform evolving with time as the intra-operative ST-elevation event took place. The main morphological feature is apparently the ST-elevation (Figure 2, supplemental video F2). The 3D image shows a visible difference before and after the event (26 beats in 0–20s vs. 40 beats in 30–60s). Binary classification using the SVM shows a perfect accuracy of 1.

The 3D image of the 66-min ISTE dataset (4,299 pulses) combined with 10-min healthy control ECG data (754 pulses) shows that the cycles from the ISTE dataset form a trajectory. This image captures the dynamical change of the ST-elevation and its gradual improvement within the 66-min period (Figure 3A & 3B, Supplemental video 3A). To quantify the time evolution of the ISTE dataset, we separate it into 6 subgroups (index ranges: 1–280, 290–580, 850–1,500, 1,500–2,300, 2,300–3,300, 3,300–4,299) to quantify the improvement in comparison with the healthy control data, and with the waveform before the ST-elevation (supplemental Table S1) The SVM analysis shows a total accuracy of 83.63 % and macro-F1 86.9%, which suggests the potential of the proposed algorithm to quantitatively measure the time-evolving dynamics.

**Nicardipine ABP Dataset**

The 3D image of one 120-s data of Nicardipine dataset comprising 187 ABP pulses shows that the 187 points form a trajectory-like structure (Figure 4A & 4B). This trajectory indicates the hemodynamic effect induced by Nicardipine. An analysis of 2,364 ABP pulses from 12 cases shows 12 trajectory-like structures corresponding to the 12 cases, which indicates the inter-individual variation. Despite this interindividual variation, these trajectories move toward in a common direction (Figure 4C&4D, supplemental video F4D), which indicates the universal impact of Nicardipine on our hemodynamic system. The SVM analysis shows a total accuracy of 95.01% and macro-Fi 96.9%.

**ETI ABP Dataset**

An analysis of one case of the ETI dataset comprising 223 pulses in 200 seconds shows that successive pulses forms a trajectory in the 3D embedding (Figure 5A & 5B, supplemental video F5B). After endotracheal intubation, the trajectory moves away from its baseline location and returns back to it after a while, forming a loop-like structure. Adding another 8 cases of endotracheal intubation (2,957 pulses in total) yields 9 loop-like trajectories (Figure 5C, supplemental video F5C), which is similar to the finding from the Nicardipine ABP Dataset. More comparison with derived features is shown in Supplementary C.4.

**Combined ABP datasets for data partitioning**

Finally, we combined the Nicardipine dataset with the ETI dataset to gain further insights into the ETI dataset. Specifically, we measured the angular readings between the ETI trajectory and the nearest Nicardipine trajectory (Figure 6). Among the 12 trajectories of the Nicardipine data, No. 3 is the nearest to ETI data in terms of the DDist. We measured the dynamic angle of the ETI trajectory with respect to No. 3 Nicardipine dataset to differentiate the ETI trajectory into two parts: the relatively parallel part (77–125th pulse) and the relatively perpendicular part (133–169th pulse). The angle of the relatively parallel part is 141.5 degree (122.9–149.1), and the angle of the relatively perpendicular part is 60.5 degree (53.3–79.0). Hypothesis test shows a statistically significant difference ($p<10^{-5}$). More combined analysis results are reported in Supplementary C.3.

# Discussion

This is the first study to investigate the cardiovascular waveform using modern unsupervised machine learning tools. This approach "learns" directly from the whole waveform instead of using the derived features. The results show that the 3D imaging provided by the DMap helps visualize the dynamics in a compact form. In addition, the analysis shows that the inner physiological dynamics can be observed on both the intra-individual and inter-individual levels. The ECG analysis shows a hierarchical structure developed from the non-ST elevation ischemic ECG waveforms and the time-evolving nature of ECG morphology during an ST-elevation. For the ABP waveform, in addition to capturing the intra-individual dynamics in response to Nicardipine, the analysis reveals a universal hemodynamic response to Nicardipine despite the influence of inter-individual variability. Furthermore, the analysis allows us to classify pulse waveform dynamics into two types: those that are related to the vasodilatory effect and those not.

**Data Characteristics Captured by Manifold Learning**

The consecutive pulse-to-pulse morphology reflects changes with time. The 3D image suggests that consecutive ECG beats or ABP pulses tend to gather together in the high dimensional space as a trajectory. The trajectory can be further quantified. In the ETI dataset, the trajectory of ABP pulses in response to the endotracheal intubation stimulus moves away and returns to the baseline state, forming a loop. The ISTE data shows a trajectory featuring an ST-elevation episode and the subsequent evolution of the ECG waveform.

The MGH/MF database focuses on the clustering effect within the scope of the non-ST-elevation ECG waveform. The ischemic ECG waveforms cluster into a hierarchical structure, representing a clinical meaningful distribution among the healthy control, ischemic ECG, and clinical unstable angina. The 3D image representing the entire ECG waveform serves as an adjunct to a visual inspection of the subtle differences among each ECG waveform. This finding coincides with the known automatic classification potential of unsupervised manifold learning.

**Clinical interpretation**

Unstable angina is a potentially life-threatening situation that exhibits a NSTEMI pattern[1,2]. Our results show some similar features between unstable angina and ischemic ECG yet differences exist. Unstable angina exhibits a more concentrated and extreme clustering of ECG waveform than ischemic ECG. Although the unsupervised method was unaware of the clinical condition labeling, the classification results are congruent with our clinical knowledge.

With MGH/MF and ISTE datasets, we demonstrate clinical values of the DMap in both ST-elevation and NSTEMI pattern. The close relationship between the ischemic ECG pattern, ischemic heart disease, and clinical outcome is well known. Moreover, the unsupervised manifold learning in this study can be used for consecutive beat-to-beat monitoring over a long period. It automatically organizes thousands of beats into a 3D image to provide a complete picture at a glance; this is convenient compared with the direct visual interpretation of the long-term ECG waveform. In addition, the ability of consuming a large quantity of ECG waveform may help eliminate the sampling error and improve the accuracy of outcome assessment.

Hedén et al. have reported that applying artificial neural network to detect acute myocardial via the ECG features derived from ST segment or T wave[18]. Other studies also have reported algorithms using features of ST segment provide diagnostic values[15,19,20]. Our results using direct waveform input are consistent with these previous studies. This fact indicates that despite being unaware of the physiological knowledge, the manifold learning approach can capture the major information inside the data, which is the ST morphological difference in the MGH/MF dataset and the evolving of ST-elevation morphology in the ISTE dataset.

Compared with the ECG signal, the ABP waveform contains information that is more relevant to the entire circulation system. The DMap visualizes the dynamics for each case in response to the endotracheal intubation procedure or the administration of anti-hypertensive medication as a trajectory in the 3D image. The relationship between the effect of antihypertensives and the derived indices of ABP has been well-established[14,21,22]. In this study, how the intravenous antihypertensive medication, Nicardipine, elicits ABP waveform change on the pulse-to-pulse scale is well captured by the DMap. That is, the trajectory captures the physiological effect in a good temporal resolution. The loop-shape trajectory in response to the endotracheal intubation suggests that the physiological status returns to the baseline status on a different route. Although there is no study particularly addressing the ABP waveform regarding the endotracheal intubation procedure, from the physiological perspective, both vasoconstriction effect and increased cardiac contractility may participate in the process. The combination of the Nicardipine and ETI datasets provides further interpretation through their geometric relationship; the data can be segmented into those related to the vasodilatory effect and those not (Figure 6). Further research is warranted to elucidate what the DMap grasps in terms of physiology and how to integrate it into clinical practice.

**Methodological Considerations**

There has been a rich knowledge about ECG and ABP waveforms. The standard interpretation of the ECG morphology relies on the visual inspection of waveform landmarks, such as P wave, QRS complex, and ST segment, and their relationship. Similarly, the augmentation pressure and augmentation index[4,6,21,23] derived from the ABP pulse waveform provide information on the cardiovascular system. Owing to the available rich information, computer algorithms for waveform analysis have been developed for clinical usage. Methods such as neural network and supervised machine learning have been employed to analyze a large amount of features derived from pulse waveforms[18,24].

The proposed manifold learning approach is different from the other approaches in several ways. First, it "learns" directly from the raw waveform, rather than the derived features. Theoretically, it may capture subtle waveform difference that is overlooked by the designed feature or by visual inspection. Second, the DMap is unsupervised; that is, the labeling information is ignored in the learning process. Hence, the learned features are intrinsic. Third, the high dimensional data is organized into a 3D image for the inner dynamic visualization. If the pulsatile data are consecutive in time, the trajectory indicates that the inner dynamics continuously evolves with time. If the data are labeled, we may observe the hierarchical tree structure from the 3D image. This visual information may help physicians grasp the whole physiological dynamic picture. Moreover, owing to its rigorous theoretical support, we do not run into the black box issues commonly encountered in other machine learning tools, such as the deep neural network.

Undoubtedly, the existing waveform features contain a large amount of accumulated knowledge regarding diagnosis using ECG and ABP waveform. It is warranted to investigate the similarity and difference between these waveform features and what the DMap provides in the future study.

**Limitations and Future Works**

There are limitations in this study. While this study provides results, our analysis may not exert the full potential of manifold learning. For each dataset, waveforms from single channel were analyzed while the data contains other channels probably providing additional information; for example, the simultaneous recorded cardiovascular waveforms, such as central venous pressure or photoplethysmography, were not considered. Another limitation is that the output information of manifold learning is not an easy-to-use 0-to-100 scale. While there is potential to visualize a long-term waveform by the DMap, the longest signal we explore in this paper is limited to 66 mins, and it is not yet clear how it performs for a longer data, like 10 hours or 14 days long. Further exploration of these limitations is critical for the clinical application.

Current results bring new opportunities in cardiovascular waveform analysis. In addition to the combination of different modalities of data, it is warranted to explore the inter-individual variability from the spectral clustering perspective, to perform prediction through the extrapolation of trajectories in time sequenced data, to integrate with known waveform features and numeric readings, and to compactly organize long period and multimodal waveforms. We forecast that manifold learning will help both gain more information and integrate information from many aspects in clinical studies.

# Sources of Funding

The work was supported by the National Science and Technology Development Fund (MOST 108-2115-M-075-001-) of Ministry of Science and Technology, Taipei, Taiwan and LEAP@Duke program of National Applied Research Laboratories, Taipei, Taiwan.

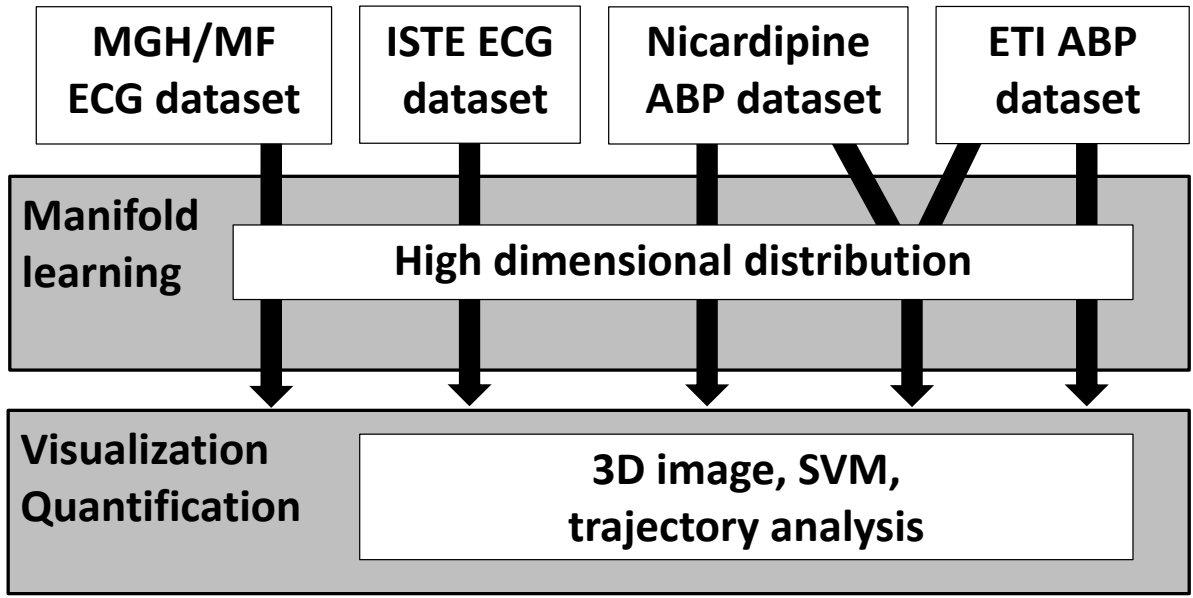

**Figure 1.** Schematic diagram of the methods

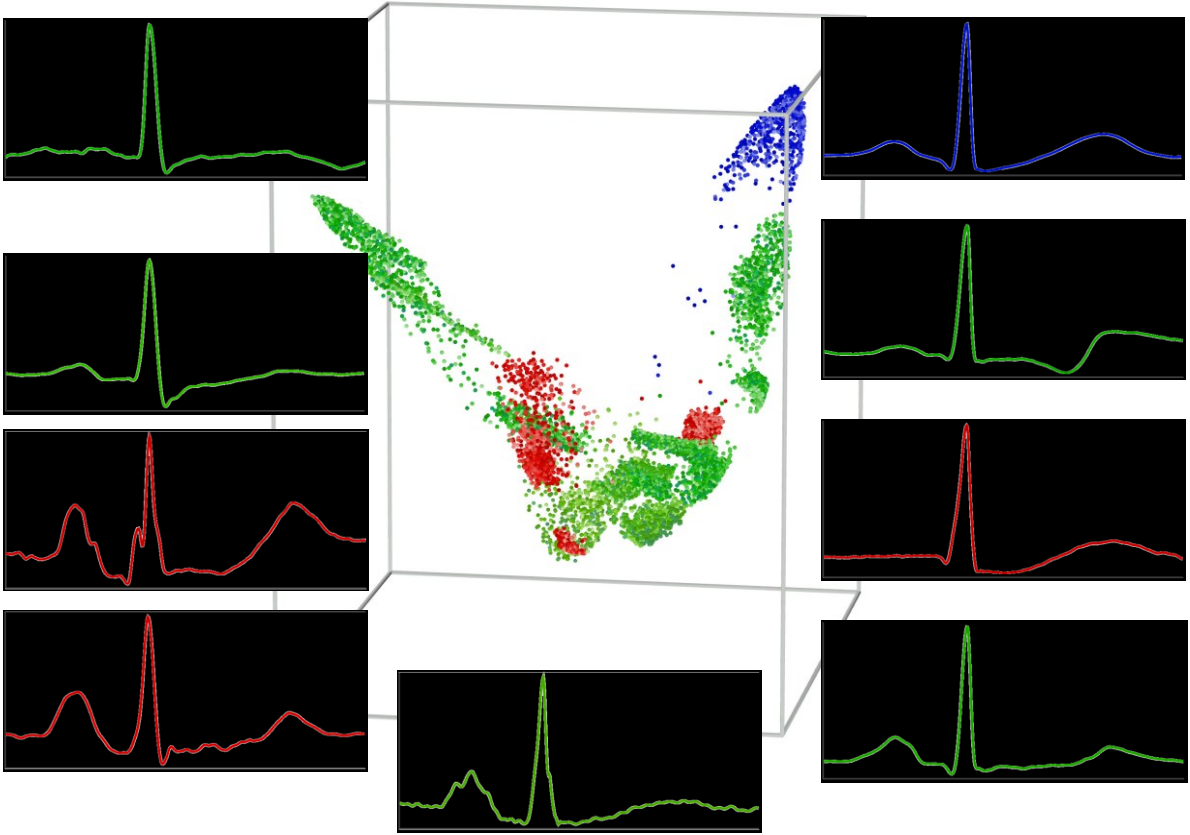

**Figure 2.** The 3D image of 10,583 electrocardiographic (ECG) beats from the MGH/MF database for non-ST elevation ECG waveform analysis. It shows that data points representing ECG beats are aggregated by their groups, forming a hierarchical tree structure. Color labels

represent groups, including clinical unstable angina (red), ischemic ECG pattern (green), and healthy control (blue). The representative waveforms are marked for comparison.

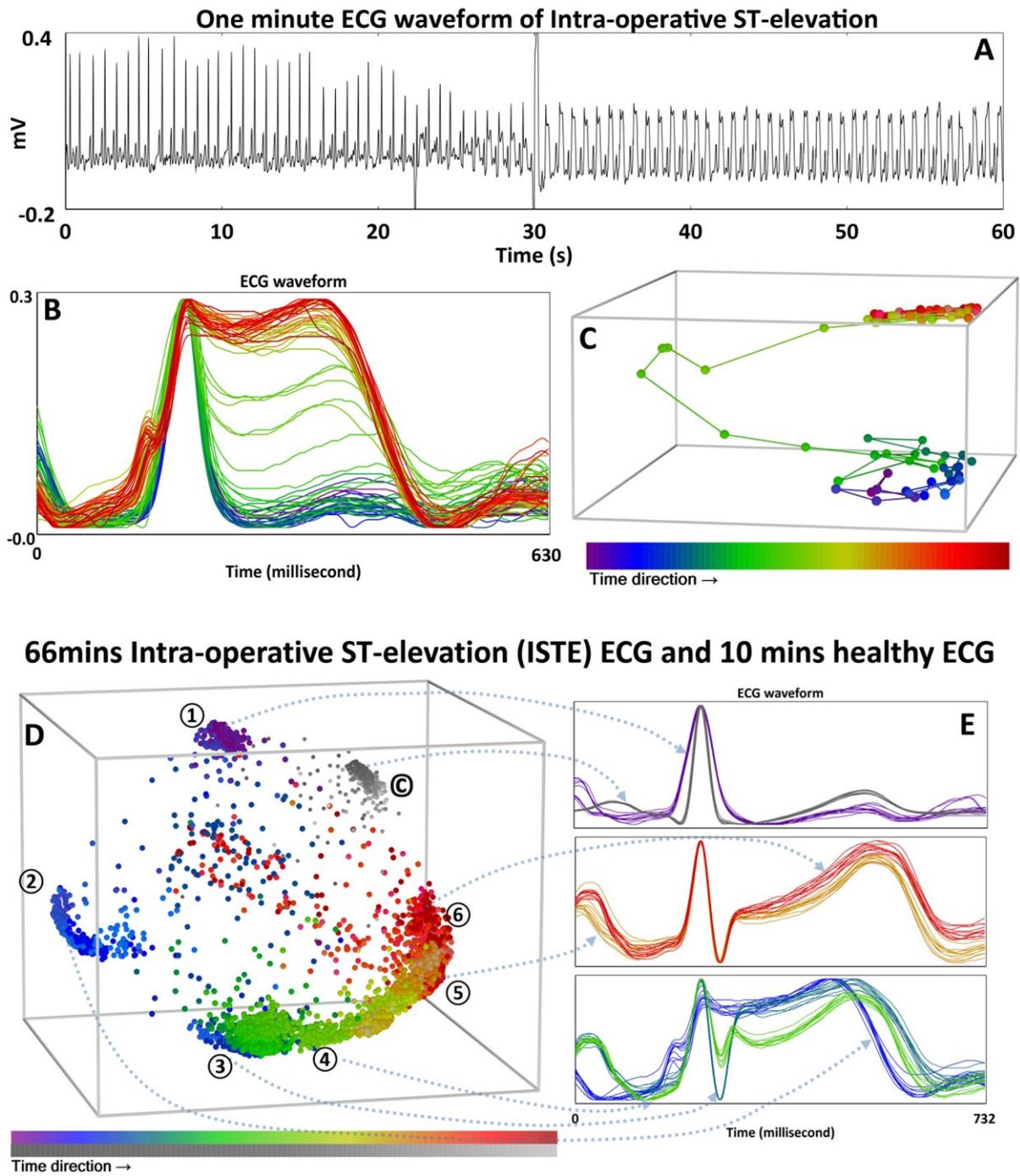

**Figure 3.** One minute electrocardiographic (ECG) data and 66 minutes ECG data recorded from an intra-operative ST-elevation event are presented. Panel A: 1-min ECG tracing shows a transition of the ST segment. Panel B: normalized successive ECG waveforms in 1-min

period with each successive beat color-labeled in the time sequence. Panel C: the 3D image with each successive beat color-labeled in the same way as that in Panel B. Panel D: the 3D image of 4,299 consecutive ECG beats from the ISTE (Intra-operative ST-Elevation) data colored in time sequences, combined with 10 minutes of healthy ECG data as the control (754 beats) in the gray color; Panel E shows the corresponding normalized ECG beats evolving with time. The one minute ECG (panel A, B, C) shows the dynamic nature of the events, while the 66-min data (panel D, E) shows the relationship between different states before the events (①), the subsequent time-evolving trajectory (②-⑥) moving toward the normal state (①), and the healthy control (©) corresponding to the improvement of clinical condition.

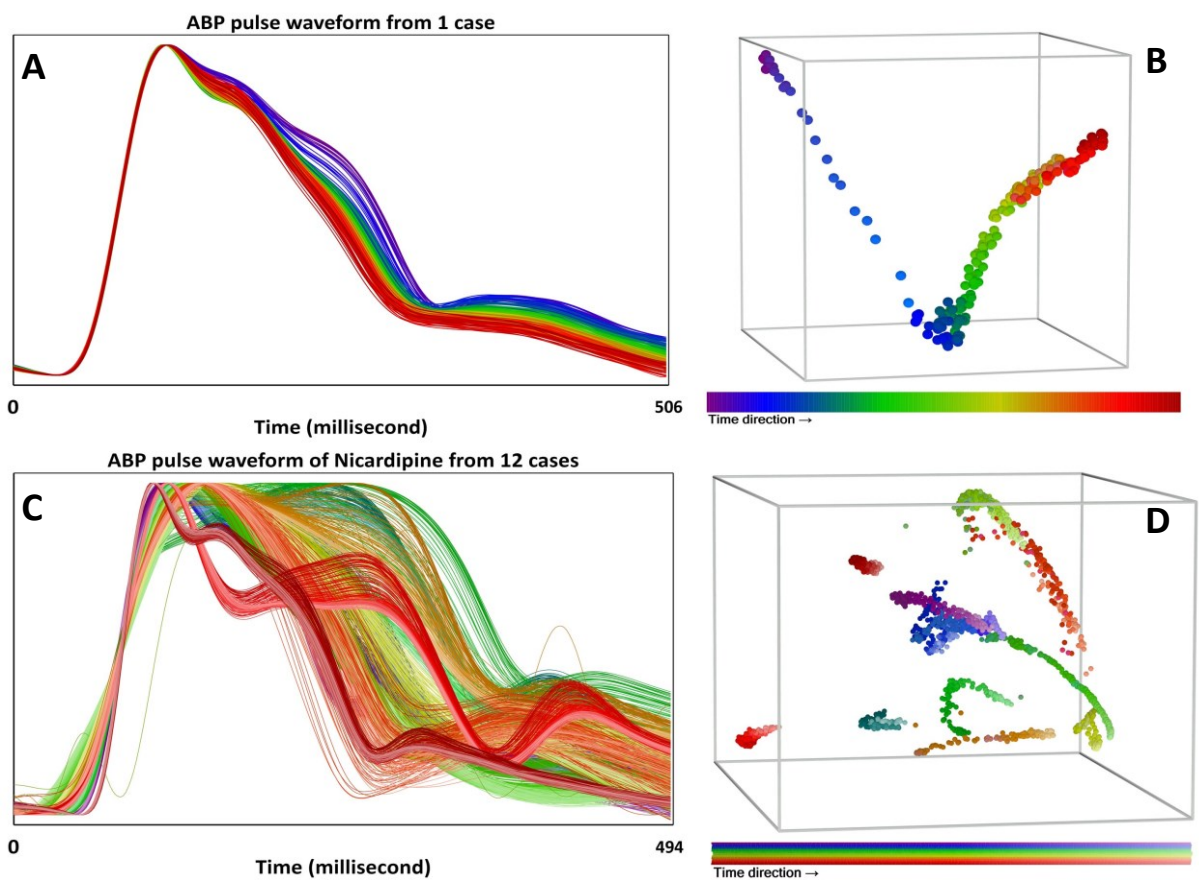

**Figure 4.** The Nicardipine effect on the ABP waveform in a single case and in 12 cases. Panel A shows 187 pulses from a 120-s segment of an ABP recorded during the intravenous bolus dose, with each successive beat color-labeled in the time sequence. Note the changes between successive pulses are subtle and difficult to interpret from the ABP waveform by naked eyes. Panel B shows the corresponding 187 pulses comprising a trajectory, also with each successive beat color-labeled in the time sequence. Panel C shows 2,303 ABP pulses from 12 cases, with pulses from different case labeled with different colors. The color is fading as time evolves. The 3D image (panel D) of 2,364 ABP pulses from 12 cases forming 12 trajectories moving toward a common place, the right lower corner. The trajectories show the inter-individual difference and the common effect of Nicardipine.

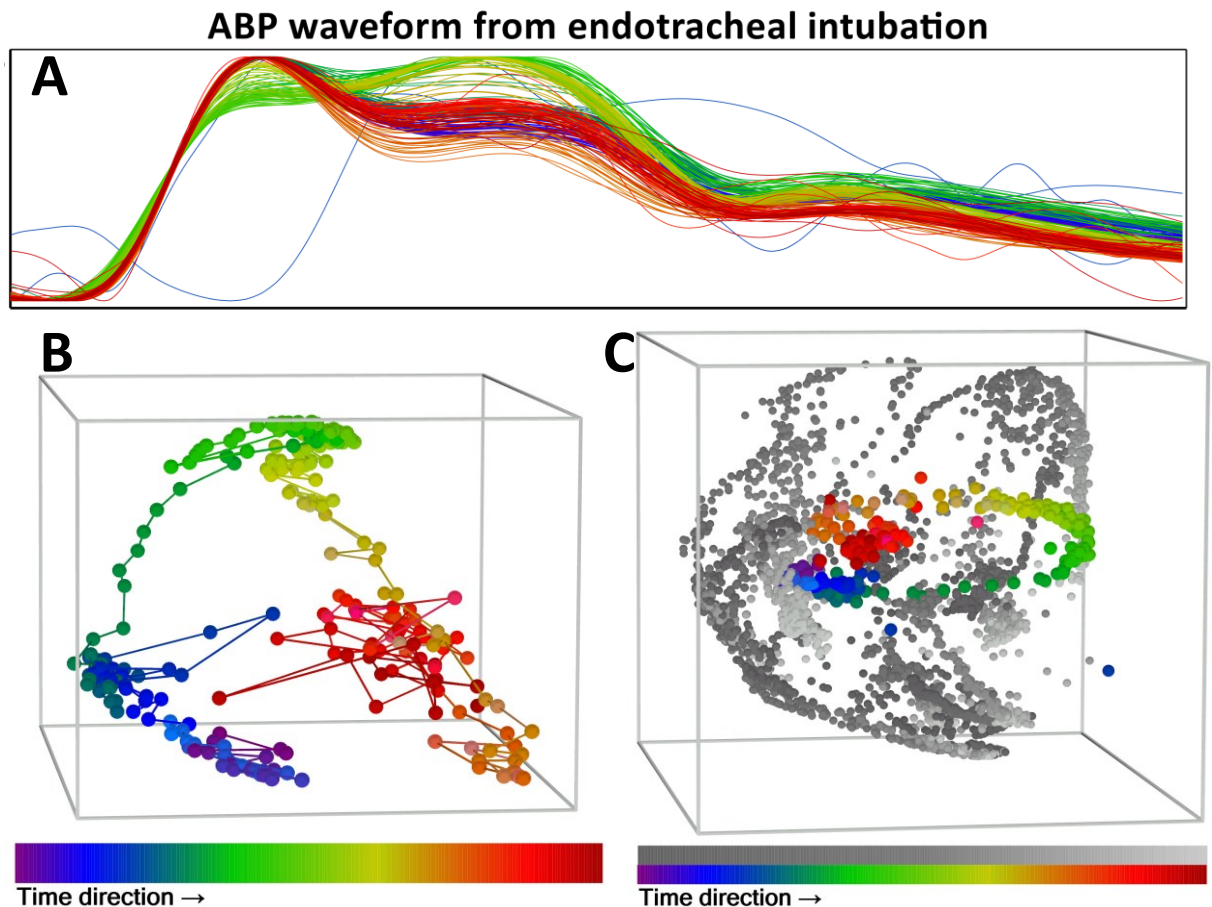

**Figure 5.** The endotracheal intubation effect on the ABP waveform. Panel A: the 223 pulses from the ABP waveform of a single case of endotracheal intubation, with each successive beat color-labeled in the time sequence. Panel B: the 3D image of the single case data, also with each successive beat color-labeled in the time sequence. Panel C: the 3D image of all pulses, including pulses from Panel A, and those pulses from another 8 cases, comprising 2,957 pulses. The pulses from another cases are colored in gray, with the fading effect indicating the time direction. The 3D images show that the trajectory in response to the endotracheal intubation moves away from its baseline location and returns back to it after a while, forming a loop-like structure.

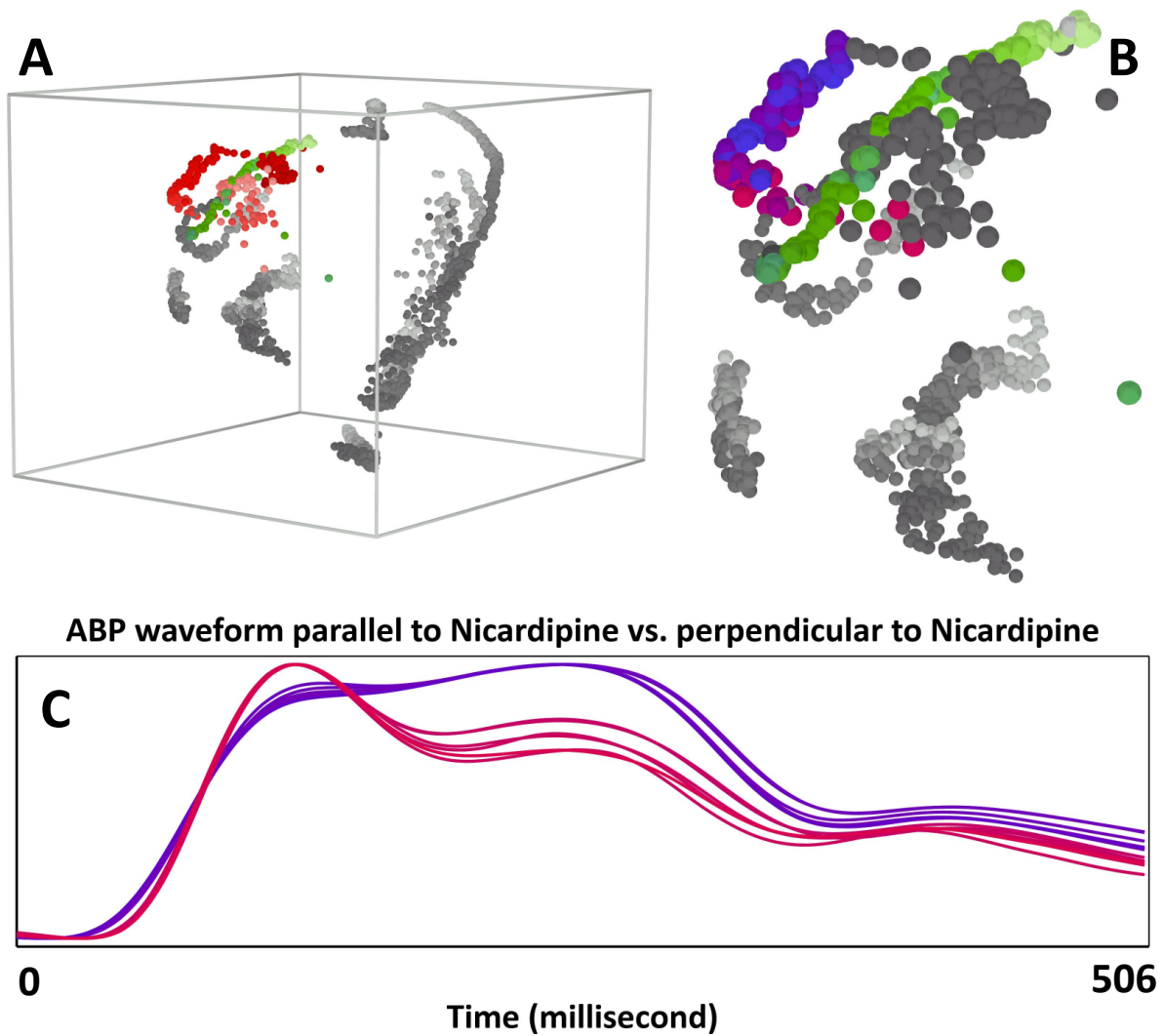

**Figure 6.** Combination of the ABP waveforms from the ETI (Endo-Tracheal Intubation) dataset and the Nicardipine datasets leads to a new way of data partitioning according to the trajectory analysis. Panel A shows the 3D embedding of the ETI dataset (red) and the case of the Nicardipine dataset (green) that is closest to the ETI dataset against other Nicardipine dataset cases (gray). Note that the trajectory of the ETI dataset forms a loop-like structure. Panel B is a zoomed in picture of Panel A, which shows that the ETI dataset forms a loop-like structure, and it is colored differently. It is colored (purple-red) according to the angle with respect to the Nicardipine dataset (green). Panel C shows the normalized ABP pulse waveforms from different sections of the loop-like structure of the ETI dataset, determined by the angle with respect to the Nicardipine dataset. Parallel direction means it is reciprocally related to the vasodilatory effect, whereas perpendicular direction means it is unrelated to the vasodilatory effect.

# Supplemental Material

## A. Technical details of manifold learning

We have seen how the three-dimensional (3D) image of a waveform signal via the diffusion map (DMap) and other quantifications work in the main article. In this supplemental material, we further describe the technical significance of the DMap and its role in machine learning.

From the aspects of artificial intelligence and machine learning, we can colloquially say that a machine is not a human expert, but it has several capabilities beyond human experts. It can faithfully follow assigned rules for the tasks and help capture patterns behind the complicated and huge amount of data; for example, playing chess or reading electrocardiography (ECG). The rules are usually designed by human experts, or learned from human experts. On the high level, we can view the output of the DMap as an "integration" of the overall (possibly) nonlinear structure of the inner dynamics we have interest. As a result, we can proceed with visualizing the waveform via the 3D image and carrying out further statistical analysis. Now we provide details.

### A.1 Model

From the model perspective, we take the mathematical object called "manifold" into account. In layman's terms, a manifold is a generalization of a smooth curve or a smooth surface in a 3D space. It can be viewed as a "high dimensional nonlinear surface". Readers having interest may consult mathematical details[1]. We model that each beat or pulse is a point lying on a manifold[2,3]. See Figure S1 as an example.

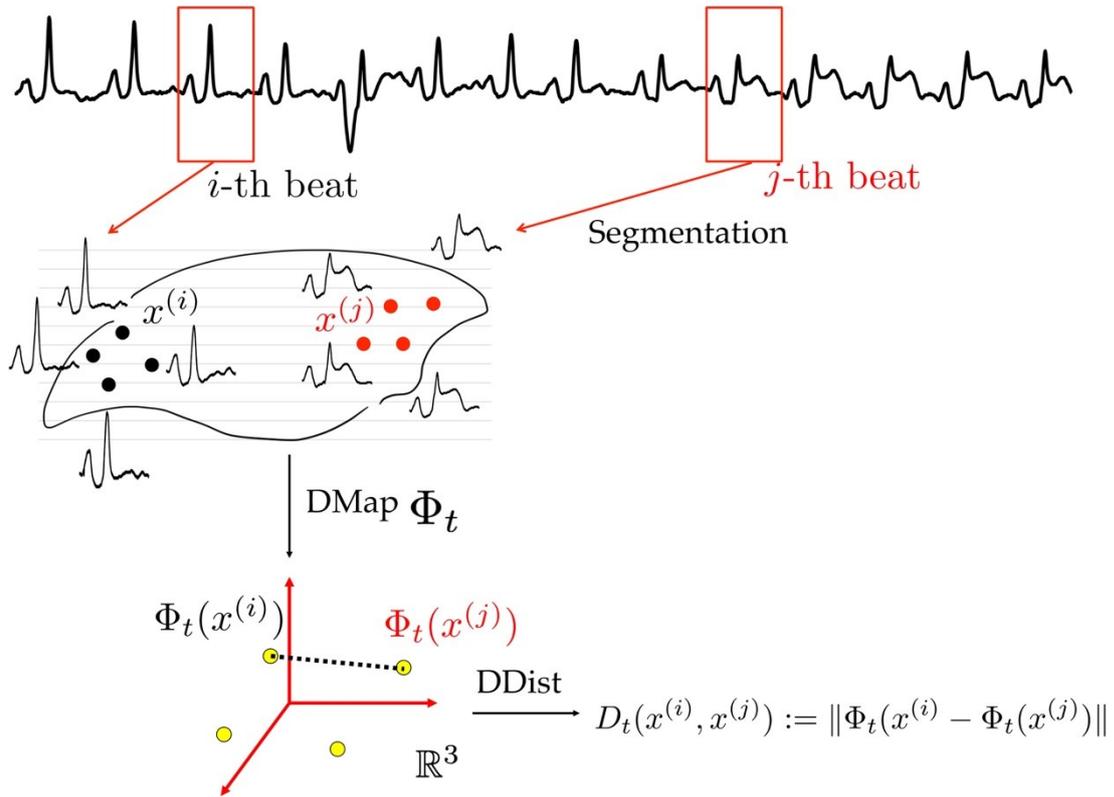

**Figure S1.** An illustration of the manifold model and the technical steps of the DMap and DDist. The ECG waveform shown in the top is from the ISTE database. In this illustration, $\hat{d} = 3$. The DDist between two pulses $x_i$ and $x_j$ is defined as the length of the dashed line between $\Phi_t(x_i)$ and $\Phi_t(x_j)$.

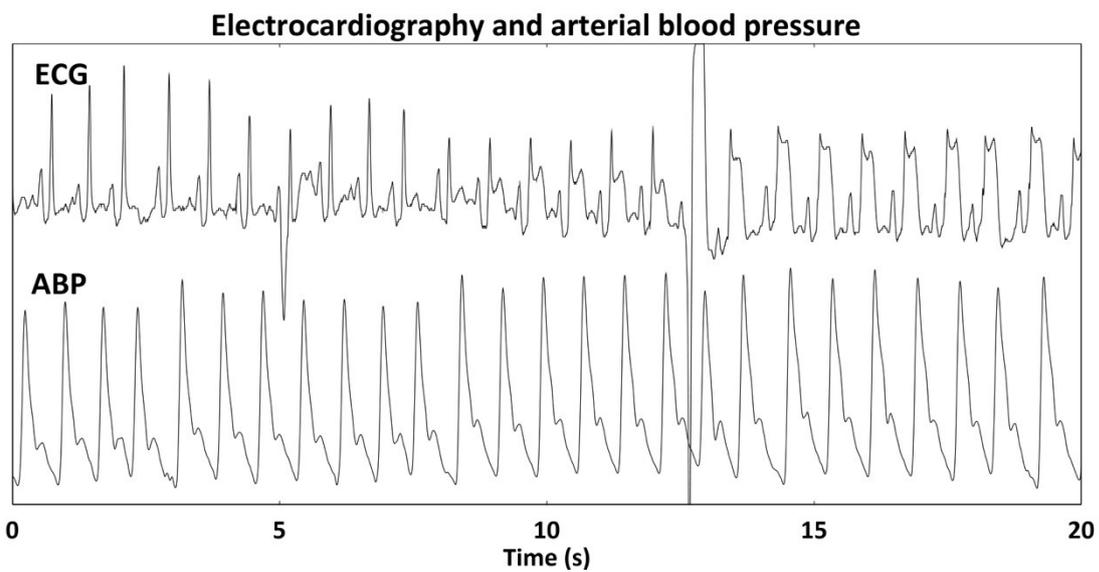

**Figure S2.** Twenty-second electrocardiographic (ECG) and arterial blood pressure (ABP) data simultaneously recorded from an episode of intraoperative ST-elevation show dynamic waveform changes.

## A.2 More about the DMap

With the manifold model, we analyze the waveform by the manifold learning tool, the DMap, to extract the complicated relationship among those pulses, and hence the inner dynamics. In layman's language, the DMap organizes dominant information shared by all pulses via the eigendecomposition. Intuitively, the eigendecomposition in the DMap is a complex computation procedure that patches microscopic information shared by consecutive pulses, and generates a macroscopic picture of the whole waveform. In general, the DMap is a nonlinear dimensional reduction method commonly considered in the machine learning society[8]. Dimensional reduction is a commonly applied statistical technique that reduces the number of parameters in the dataset (reduce the dimension of the original data) so that the contained information is preserved or similar between the original dataset and the dimensional reduced dataset.[4] For readers not familiar with this kind of nonlinear method, it can be understood as a nonlinear generalization of the famous linear method, the principal component analysis (PCA).[5] Researchers usually use PCA to find statistically linearly correlated factors.[6]

The output of the DMap in general is of lower dimension compared with the original dataset, but is still of dimension higher than three. Intuitively, we view each dimension of the data as a separate scalar variable, and can treat them and a multivariable statistical quantity. In this work, the dimension of the input database to the DMP, that is, the set of beats or pulses from the waveform, depends on the sampling rate.

## A.3 3D image visualization

The output of DMap can be converted to a 3D image for the visualization purpose. See Figure S1 for an illustration of converting a waveform into a 3D image. In the 3D image, each pulse is represented as a point, and the whole waveform is represented as a set of point cloud, which "shape" integrates the nonlinear information hidden in the waveform. We have seen that the 3D visualization is a useful feature of the DMap. With the 3D image, we can take advantage of the ability of human vision and brain to identify various shapes and establish the structure. As the raw waveform may contain nuances information, and the useful information might easily be overlooked by naked eyes, particularly when the waveform is long, we may grasp the whole picture at a glance via the 3D visualization of the waveform.

## A.4 Analysis via the DDist

However, in general, the 3D visualization only shows partial information, and the output of the DMap contains additional information. The diffusion distance (DDist)[3] is a tool that measures the similarity between two pulses by calculating their distance in the output of DMap. Based on the developed theory[3,7], a smaller DDist indicates the more similarity between two pulses. See Figure S1 for an illustration of how the DDist is calculated. In this study, the result of the DMap is visualized by using 3D imaging, and the quantification is based on the DDist.

A technical detail of DMap and DDist should be mentioned. There are many distances we can choose to quantify similarity between two pulses. We choose the DDist for two purposes. First, it is a natural combination, both in practice and in theory, with the DMap. Second, since clinical waveform data is usually not clean, and the denoising step is usually needed before evaluating the distance. However, the robustness property of DMap and DDist could directly handle this noisy issue.[8] In other words, the DMap and DDist combine seamlessly the needs for denoise and evaluating the distance.

## B. Numerical implementation details

In this section, we provide numerical implementation details with the mathematical formula for the reproducibility purpose. The code can be downloaded from https://hautiengwu.wordpress.com/code/

### B.1 Data preprocessing

To analyze the pulse waveform of an ABP signal $A(t)$ (or similarly for an ECG signal), we break it into segments so that each segment $x^{(i)}$ is the $i$-th ABP pulse containing one pulse waveform cycle. See Figure S1 for an illustration of this segmentation procedure.

We normalize the $x^{(i)}$ to separate the blood pressure information from the wave shape morphology, and denote the result as $\bar{x}^{(i)}$. Suppose there are $N$ consecutive ABP pulses from $A(t)$. By the above procedure, we have a set of normalized pulses (or usually called the point cloud), and we express this set as

$$\chi := \{\bar{x}^{(i)}\}_{i=1}^{N} \subset L^2([0, L]),$$

where $L$ represents the length of each pulse, determined as their minimal length. Since the waveform data is usually contaminated by noise, $\chi$ is a ``noisy'' collection of ABP pulses. We mention that we can view each entry of $\bar{x}^{(i)}$ as a ``feature'' associated with the $i$-th pulse, while it is more common to read the whole pulse or derive features from $\bar{x}^{(i)}$.

**B.2 DMap**

The Diffusion map is one of the manifold learning algorithms undergoing rapid development, both in theory and practical aspects. It is characterized by solid mathematical and statistical supports.[3,7,8] It handles the collection of pulses as a point cloud $\chi$ to construct a $n \times n$ affinity matrix $W$ so that

$$W_{ij} = e^{-d|x_i - x_j|^2/\varepsilon}, \text{ for } i,j = 1, \ldots, n,$$

where the bandwidth $\epsilon > 0$ is chosen by the user. The $W_{ij}$ contains the difference information between each-two-point. In this study, we standardized the $\epsilon$ as the 25% distribution of each-two-point distances. The *degree matrix* $D$ of size $n \times n$ was constructed by

$$D(i,i) = \sum_{j=1}^{n} W(i,j), \text{ for } i = 1, \ldots, n.$$

With matrices $W$ and $D$, a random walk on the point cloud $\chi$ is represented by the transition matrix $A$ as:

$$A \coloneqq D^{-1}W.$$

Then, we perform eigen-decomposition $A = U\Lambda V^T$, so that $1 = \lambda_1 > \lambda_2 \geq \lambda_3 \geq \cdots \lambda_n$ are eigenvalues of $A$, and $\phi_1, \phi_2, \ldots, \phi_n \in \mathbb{R}^n$ are the associated eigenvectors. With this eigen-decomposition, the DM is defined as

$$\Phi_t : x_j \mapsto \left( \lambda_2^t \phi_2(j), \lambda_3^t \phi_3(j), \ldots, \lambda_{\hat{d}+1}^t \phi_{\hat{d}+1}(j) \right),$$

where $j = 1, \ldots, n, t > 0$ is the diffusion time chosen by the user, $\hat{d}$ is the embedding dimension chosen by the user. Note that $\lambda_1$ and $\phi_1$ are discarded because they contain no information.

In this study, the eigenvalues associated with the pulse waveform data generally decay fast so the eigenvalue of $\hat{d} > 15$ is small enough and can be neglected. We thus take $\hat{d} = 15$ in the analysis. As a result, the *ith* pulse, $x_i$, is mapped to a 15-dim space; that is, the output of the DMap is composed of 15 parameters.

Based on the output from DMap, the *diffusion distance* (DDist) can be obtained as the length of the straight line connecting $\Phi_t(x_i)$ and $\Phi_t(x_j)$; that is, the L² distance between $\Phi_t(x_i)$ and $\Phi_t(x_j)$:

$$D_t(x_i, x_j) := \|\Phi_t(x_i) - \Phi_t(x_j)\|_{l^2}.$$

Clearly, the DDist is a scalar value assigned to each pair of points (pulses). In this study, we use the DDist to measure the similarity. We also use it as the yardstick to perform null hypothesis test between two groups.

Further analysis, such as profiling the trajectory, is based on the output of the DMap as well. We mention that when the DDist was used to evaluate the difference between two groups, we first calculated the mean value of each groups as the "geometric center" and then measured the DDist between these two geometric centers. Another possible method is to calculate the average value from all possible pairs between two groups. In this study, what we have seen from the 3D image of DMap shows clustering distribution of different groups, so we chose the current measurement of group distance.

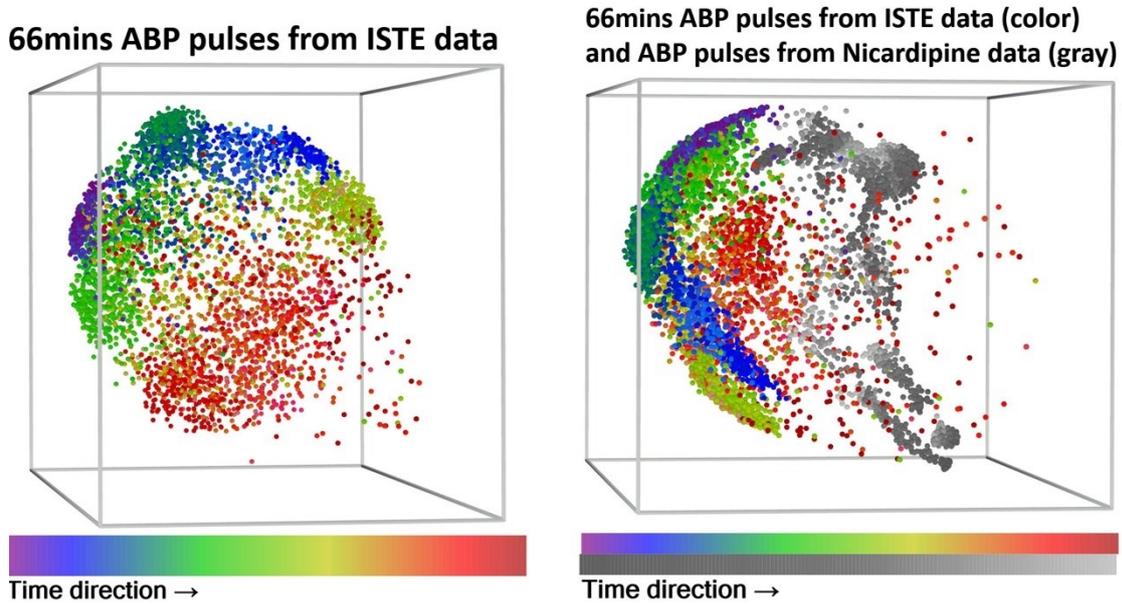

Figure S3. Top panel: The bottom left panel shows the 3D image of the ABP waveform, and the bottom right panel shows the 3D image of all pulses from the ABP waveforms from the ISTE database and the Nicardipine database.

## C. More analysis results

## C.1 More statistical results of dynamics separation

We provide the confusion matrices, sensitives and precisions of each group for the SVM analysis of the NSTEMI ECG dataset, the ISTE ECG dataset, and the Nicardipine ABP dataset in Tables 1, 2 and 3.

We further report a sensitivity analysis for the dynamics separation. We vary the following factors to study their impacts: the dimension of the DMap output, handling the imbalance data or not, applying different kernels in the SVM. The result is shown in Table S4. The sensitivity test results are similar in all situations and support the results.

**Table S1.** Diffusion distances between each group in the ISTE dataset to the healthy control and to Group 1. The mean values and 95 % confidence intervals are expressed in an arbitrary unit with the scale $10^{-3}$.

|  | Group 1 | Group 2 | Group 3 | Group 4 | Group 5 | Group 6 |
|---|---|---|---|---|---|---|
| **Distance to healthy control** | 3.35 (3.33–3.38) | 8.58 (8.42–8.74) | 9.18 (9.02–9.34) | 9.29 (9.22–9.34) | 8.85 (8.79–8.91) | 7.82 (7.72–7.92) |
| **Distance to Group 1** |  | 8.04 (7.86–8.21) | 9.44 (9.27–9.60) | 9.63 (9.59–9.68) | 9.61 (9.56–9.66) | 8.59 (8.49–8.69) |

**Table S2.** The confusion matrix of NSTEMI ECG database when we applied the SVM.

|  | Ture label |  |  | Precision |
|---|---|---|---|---|
| **Prediction** | 692 | 32 | 0 | 95.6% |
|  | 0 | 8246 | 1 | 100% |
|  | 0 | 286 | 1326 | 82.3% |
| **Recall** | 100% | 96.3% | 99.9% |  |

**Table S3.** The confusion matrix of ISTE ECG database when we applied the SVM.

|  | True label |  |  |  |  |  | Precision |
|---|---|---|---|---|---|---|---|
| **Prediction** | 274 | 4 | 0 | 0 | 2 | 0 | 97.9% |
|  | 4 | 279 | 3 | 1 | 0 | 3 | 96.2% |
|  | 1 | 35 | 533 | 11 | 10 | 60 | 82% |
|  | 0 | 0 | 24 | 612 | 164 | 0 | 76.5% |
|  | 0 | 1 | 0 | 29 | 760 | 210 | 76% |
|  | 0 | 6 | 27 | 0 | 133 | 834 | 83.4% |
| **Recall** | 98.2% | 85.9% | 90.8% | 93.7% | 71.1% | 75.3% |  |

**Table S4.** The confusion matrix of the Nicardipine ABP dataset when we applied the SVM.

|  | True label |  |  |  |  |  |  |  |  |  |  |  | Precision |
|---|---|---|---|---|---|---|---|---|---|---|---|---|---|
| **Prediction** | 110 | 0 | 0 | 0 | 0 | 0 | 0 | 0 | 0 | 0 | 0 | 0 | 100% |
|  | 0 | 120 | 0 | 0 | 0 | 0 | 0 | 0 | 0 | 0 | 0 | 0 | 100% |
|  | 0 | 0 | 165 | 0 | 0 | 0 | 0 | 22 | 0 | 0 | 0 | 0 | 88.2% |
|  | 0 | 0 | 0 | 147 | 0 | 0 | 0 | 0 | 0 | 0 | 0 | 0 | 100% |
|  | 0 | 0 | 0 | 0 | 131 | 0 | 0 | 0 | 0 | 0 | 0 | 0 | 100% |
|  | 0 | 0 | 0 | 0 | 0 | 126 | 0 | 0 | 0 | 0 | 0 | 0 | 100% |
|  | 0 | 0 | 0 | 0 | 0 | 0 | 382 | 42 | 0 | 0 | 0 | 0 | 90.1% |
|  | 0 | 0 | 0 | 0 | 0 | 0 | 0 | 533 | 0 | 0 | 5 | 0 | 99.1% |
|  | 0 | 0 | 0 | 0 | 0 | 0 | 0 | 0 | 150 | 0 | 0 | 0 | 100% |
|  | 0 | 0 | 0 | 0 | 0 | 0 | 0 | 21 | 0 | 106 | 0 | 0 | 83.5% |
|  | 0 | 0 | 0 | 0 | 0 | 0 | 0 | 13 | 0 | 0 | 152 | 0 | 92.1% |
|  | 0 | 0 | 0 | 0 | 0 | 0 | 0 | 0 | 0 | 0 | 0 | 139 | 100% |
| **Recall** | 100% | 100% | 100% | 100% | 100% | 100% | 100% | 84.5% | 100% | 100% | 96.8% | 100% |  |

**Table S5.** The sensitivity test result for the NSTEMI ECG database. The first row indicates the factor we vary in the sensitivity analysis. ACC = accuracy, MF1 = macro-F1, RBF = radial basis function.

|     | $\hat{d} = 12$ | $\hat{d} = 18$ | No SMOTE | RBF kernel |
| --- | --- | --- | --- | --- |
| ACC | 97.3% | 96.9% | 96.9% | 96.5% |
| MF1 | 96% | 95.3% | 95.1% | 94.1% |

**Table S6.** The sensitivity test result for the ISTE ECG database. The first row indicates the factor we vary in the sensitivity analysis. ACC = accuracy, MF1 = macro-F1, RBF = radial basis function.

|     | $\hat{d} = 12$ | $\hat{d} = 18$ | No SMOTE | RBF kernel |
| --- | --- | --- | --- | --- |
| ACC | 84% | 84.6% | 84.2% | 84.3% |
| MF1 | 87.4% | 87.8% | 87.2% | 87.4% |

**Table S7.** The sensitivity test result for the Nicardipine ABP dataset. The first row indicates the factor we vary in the sensitivity analysis. ACC = accuracy, MF1 = macro-F1, RBF = radial basis function.

|     | $\hat{d} = 12$ | $\hat{d} = 18$ | **No SMOTE** | **RBF kernel** |
| --- | --- | --- | --- | --- |
| **ACC** | 93.7% | 96.6% | 95.4% | 98% |
| **MF1** | 95.4% | 97.9% | 97% | 97.9% |

**C.2 The ABP waveform analysis for the ISTE dataset.**

We now show the ABP waveform analysis from the ISTE dataset. While the ST-elevation in the ISTE dataset can be easily visualized in the ECG waveform shown in Figure S2 (as well as the 3D image shown in Figure 3D), it is less clear by reading the ABP waveform. To check if such hemodynamic information also exists in the ABP signal, we analyze the corresponding ABP wave from the 66-min ISTE dataset. The 3D image of 4,420 ABP pulses (Figure S2, bottom left subplot) shows a trajectory similar to the ECG waveform, and this exhibits the ABP pulses evolving with time along the ST-elevation. This finding suggests the potential to "read the ABP waveform details" directly via the manifold learning.

**C.3 Combination of ABP waveforms from the Nicardipine and ISTE datasets.**

We combined ABP waveforms from the Nicardipine dataset with the ISTE dataset to further evaluate the amount of differences between these two datasets. The 3D image of 6,723 pulses (Figure S2, bottom right, and supplemental video F3D) demonstrates a distinct localization between the Nicardipine and ISTE datasets. An SVM analysis of these two datasets shows a total accuracy of 99.6 %.

**C.4 Comparison with traditional features**

In Figure S4, we show the 3D image of an ABP waveform from one case of the ETI dataset. The 3D images are colored by different information. On the left panel, the color indicates the time sequence, while on the right panel, the color indicates the heart rate, which is the most basic feature we can extract from the ABP waveform. The arrow indicates the start of the intubation eliciting the dynamic response of waveform morphology and the heart rate. It is clear that right after the noxious stimulation, the trajectory shifts away from the baseline, while the heart rate changes several beats after. This latency indicates that the information contained in the 3D image is different from that of the heart rate.

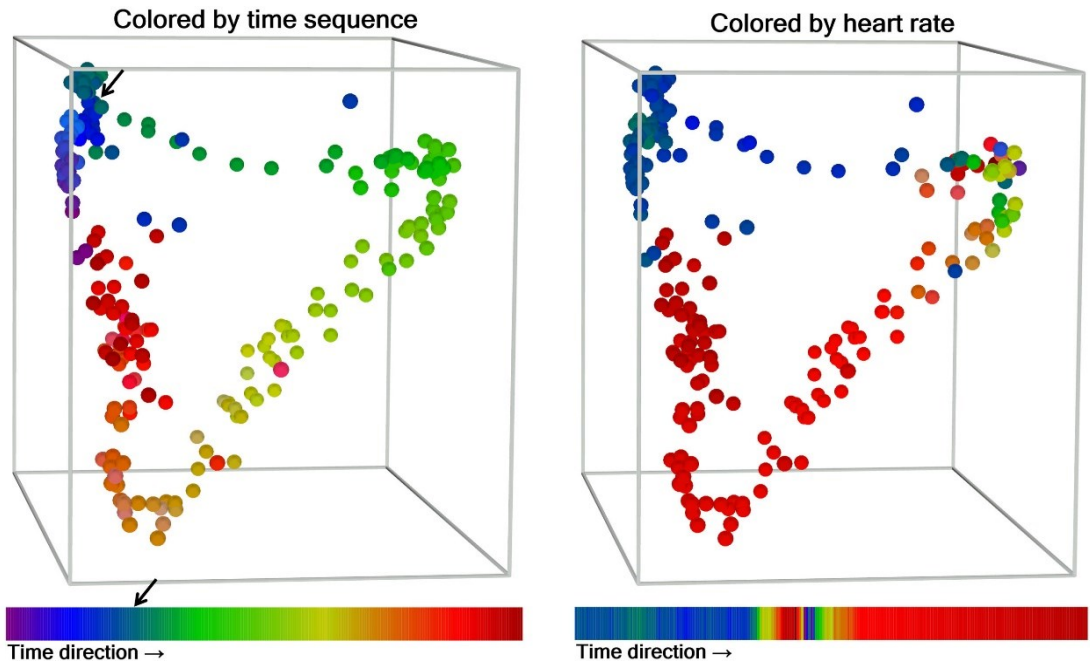

Figure S4. Arterial blood pressure (ABP) waveforms from one case of endotracheal intubation (ETI) dataset are presented as 3D images. Left panel is the color labeled by time sequence while right panel is the same data color labeled by heart rate. Arrow indicates the start of the intubation eliciting the dynamic response of waveform morphology and the heart rate. Comparing both 3D images reveals that the trajectory contains information different from the heart rate.

## D. Supplemental Video

**Video F2**
Corresponding to Figure 2, it shows the 3D animation of the MGH/MF dataset comprising 10,583 ECG beats, including clinical unstable angina (red), ischemic ECG pattern (green), and healthy control (blue).

**Video F3C**
Corresponding to Figure 3 panel C, it shows the 3D animation of 1-min ECG beats during the intra-operative ST-elevation event.

**Video F3D**
Corresponding to Figure 3 panel D, it shows the 3D animation of 4,299 consecutive ECG

beats from the ISTE (Intra-operative ST-Elevation) data colored with time sequences, combined with 10 minutes of healthy ECG data as the control (754 beats) in gray color.

**Video F4D**

Corresponding to Figure 4 panel D, it shows the 3D animation of the Nicardipine effect on 2,364 ABP pulses from 12 cases that form 12 trajectories moving upward. All cases are labeled with different colors with fading as time evolves.

**Video F5B**

Corresponding to Figure 5 panel B, it shows the 3D animation of 223 pulses from the ABP waveform of a single case of endotracheal intubation.

**Video F5C**

Corresponding to Figure 5 panel C, it shows the 3D animation of the ABP waveforms (labeled in color) of the case shown in Figure 5 Panel B, as well as another 8 cases (labeled in gray) from the ETI (Endo-Tracheal Intubation) dataset that comprises 2,957 pulses.

**Video FS3**

Corresponding to the supplementary material Figure S3 right panel, it shows the 3D animation of the combination of the 66 mins ABP waveform from the ISTE ABP dataset (color) and the Nicardipine dataset (gray).

# References for the supplementary